\pdfoutput=1
\documentclass{article}
\usepackage{amsmath}
\usepackage[utf8]{inputenc}
\usepackage{hyperref}
\hypersetup{
    colorlinks=true,
}
\title{Carbon accounting in the Cloud: \linebreak a methodology for allocating emissions across data center users}
\author{Ian Schneider\thanks{Google, Data Scientist - Research}, Taylor Mattia\footnotemark[1] \thanks{Thank you to our colleagues at Google who helped shape and improve this paper and the data pipelines and software system it describes. Thank you especially to Thomas Olavson, Savannah Goodman, Maud Texier, and Parthasarathy Ranganathan, who helped shape the direction of the paper and provided critical feedback to improve our drafts. Thank you also to Cynthia Wu and the Cloud Carbon Footprint team who supported this work and drafted the first public descriptions of the methodology described in detail here. Thank you to David Lo, Arif Merchant, Chrissy Patterson, Vincent Poncet and Cooper Elsworth for feedback on the draft. Thank you to Brock Taute, Tim Huang, Shannon Lane, Luka Rekhviashvili, and others who helped improve the software implementation of the data pipeline described in this paper.}}
\date{June 2024}

\usepackage[numbers]{natbib}
\setcitestyle{numbers}
\usepackage{graphicx}
\usepackage{float}
\usepackage{xcolor}
\usepackage{pdflscape}
\usepackage{multirow}

\newcommand{\reva}[1]{#1} 
\newcommand{\revb}[1]{#1}

\begin{document}

\maketitle

\section{Introduction}

Google has undertaken considerable efforts to reduce electricity consumption and the associated greenhouse gas (GHG) emissions from its electricity use. By 2022, Google delivered approximately three times as much computing power with the same amount of electrical power as it did five years prior \citep{googleER}.\footnote{According to Google’s platform-neutral measurement for central processing unit (CPU) resources analyzed over a five-year period.} Google uses 5.5 times less overhead energy for every unit of information-technology (IT) equipment energy, compared to the industry average \citep{googleER}. Even with these dramatic improvements in efficiency, Google consumed 22 TWh of electricity in 2022, with the majority of its electricity consumption coming from data center operations \citep{googleER}.

To reach its sustainability goals, NetZero \citep{NetZero} and 24/7 carbon-free energy (CFE) \citep{Google24x7}, and to better serve its customers, Google has embarked on a substantial effort to better understand its data center electricity use and carbon footprint. While Google has been carefully measuring its aggregate operational electricity use for a decade, starting in 2020 it began a new effort to develop granular data on hourly electricity consumption and carbon emissions by product and by location. This information helps Google better target efforts for data center efficiency and flexibility \citep{radovanovic2022carbon, KoningsteinCarbonAwareBlogPost}. It allows Google to provide carbon reporting data for enterprise customers of multiple Google products, including \href{https://cloud.google.com/carbon-footprint?hl=en}{Google Cloud} and \href{https://support.google.com/a/answer/13761003?hl=en}{Workspace}. It also allows Google to better quantify the carbon emissions of user products, like Maps or Meet. 

This paper outlines the approach that Google developed to quantify location-based emissions of its individual products. There are five primary areas of complexity. First, Google data centers have millions of different devices, including hundreds of distinct component and machine types, consuming electricity all over the world. We need to measure the electricity consumption of as many of these devices as possible and fill in the gaps when electricity measurements are not reliably available. Second, Google's ``warehouse-scale computer'' approach means that many data center resources are shared across teams and products \citep{barroso2019datacenter}. This drives higher efficiency, but presents accounting challenges. Third, Google creates many products and services that are used not only by external users, but also by company employees and internal teams. This circularity needs to be accounted for in an honest allocation of emissions. Fourth, Google consumes electricity all over the world, so we need reliable and comprehensive mechanisms for estimating carbon emissions in many different locations, many of which do not have reporting standards sufficient for our granular data requirements. \reva{Fifth, we have physical data that correlates with energy consumption in many but not all domains. For some calculations we need to fall back on allocation of energy and emissions by economic factors instead of using preferred physical factors, in line with the the Greenhouse Gas Protocol's recommendation \citep{ghgprotocol}.}

\section{Literature and State-of-the-art}

Data center electricity consumption and related emissions make up a meaningful share of global totals. In 2022, data centers accounted for 2\% of global electricity demand, and the electricity demand from data centers is forecasted to double by 2026 \citep{iea_electricity}. 

A critical way to reduce data center emissions is through more accurate accounting and improved disclosure \citep{lariviere2016better}, to enable users--especially large business users due to their scale--to make their own decisions to reduce emissions. Says \citep{gupta2021chasing}, ``although many organizations
publicly report their carbon emissions, improved accounting
(e.g., broader participation as well as standardized accounting
and disclosures) will provide further guidance on tackling
salient challenges in realizing environmentally sustainable
systems.'' 

In addition to and alongside the climate benefits,  better accounting and disclosure enable higher-quality corporate reporting of carbon emissions. Existing research summarizes corporate carbon reporting \citep{he2022corporate}. For a Cloud-provider like Google, accurate accounting and disclosure meet essential needs of major customers. These customers want (or, soon, could be required) to report their carbon emissions, and many are interested in taking steps to reduce their Cloud-related data center emissions. 

Customers and innovators are increasingly demanding accurate accounting of their own carbon footprint from their use of Cloud and other digital technologies. When a customer uses Cloud-based services like Google Cloud Platform, all of the Google Cloud emissions relevant to serving their needs fall under the customer's Scope 3 reporting. For more information on Scope 3 reporting by sector, see \citep{cdpScope3}. Research shows that historically, most companies have not completely reported Scope 3 in their CDP (formerly the Carbon Disclosure Project) Reporting
\citep{brander2023corporate}. However, this is changing as companies seek to lead in this sector, and, increasingly, when companies are required by regulation. As of March 2024, we saw that so many of our Google Cloud customers monitor and review their emissions, that over 50\% of carbon emissions are already being monitored. Disclosure is also important for growing technologies like machine learning; researchers have suggested \citep{schwartz2020green} including efficiency as a specific evaluation criteria of machine learning models.  

The approach that we developed and present here uses a mix of estimation, energy profiling, and internal resource and cost accounting. Prior research used workload proxies or measured workload duration to estimate energy consumption and carbon footprint of datacenter workloads like ML training \citep{strubell2019energy, dodge2022measuring}. However, estimated energy consumption can differ substantially from direct measurements, \citep{patterson2021carbon} and thus direct measurements of energy consumption (as we use in \citep{patterson2021carbon} and here) are much preferred when available. At the organizational level, direct measurements are not sufficient on their own because datacenter machines and workloads \revb{can be} shared amongst multiple users in a modern company and Cloud provider like Google. The primary contribution of this work is a clear methodology for allocating energy consumption to multiple users of shared data center machines, infrastructure, and software.


Besides improvements in accounting and measurement, another important area of the literature focuses on efforts to reduce the energy consumption and carbon footprint of data centers. \citep{katal2023energy} summarizes industry energy efficiency efforts and \citep{barroso2019datacenter} provides helpful information on Google-specific efforts to improve energy efficiency. Google has also worked on some specific new technologies to reduce the carbon footprint of data center compute, such as carbon-aware computing which shifts loads to times and data center locations where more carbon-free energy is available from the grid \citep{radovanovic2022carbon}. Research by Google shows how efficiency best practices and carbon-free energy procurement have helped dramatically reduce the carbon emissions associated with training machine learning models \citep{liang2022carbon}. Our hope is that improved accounting and estimation of carbon emissions can help identify key opportunity areas and catalyze new innovation to further reduce data center energy consumption and carbon emissions.



\section{Methodology}

This section describes the main contribution of this paper--a detailed description of the methodology used for granular product-level energy allocation and location-based carbon emissions accounting at Google. This accounting methodology forms the backbone of our Google Cloud Carbon Footprint reporting product \citep{GCPmethodology} and allows us to estimate the carbon emissions associated with many of Google's products and services. 

\reva{The methodology uses a mix of physical allocation, when available, and economic allocation when physical data or factors are not available. The Greenhouse Gas Protocol's Scope 3 Standard \citep{ghgprotocol}, which provides guidelines for corporate carbon reporting, says to ``consider physical allocation,'' but that if physical data is not available or is not cost-effective to gather, to ``allocate using economic factors'' \citep{ghgprotocol}. While building the Google Cloud Carbon Footprint product, we identified several cases where physical data is not available, is too expensive to collect, or where we have been unable to evaluate and verify the connection between available physical data and emissions. In these cases, we use economic allocation instead of physical allocation. This paper describes the current state, aligned with the Greenhouse Gas Protocol, where we use physical allocation when physical data is available and allocation by economic factors when physical data is not available. If more physical data becomes available, we will continue to improve the product over time by switching to allocation using physical data when possible. This will improve the accuracy of our Carbon Footprint data and help make it more actionable.}

First, we define some key concepts used throughout the section. Then, we explain the methodology boundary and exclusions. Next, we explain the allocation methodology used for different machine types and components of power usage. Then, we describe the reallocation that handles internal use of shared Google services. Next, we explain how we incorporate carbon emissions measurements. Finally, we explain how we allocate carbon emissions to Google Cloud customers for the Google Cloud Carbon Footprint reporting platform.

\label{sec: methodology}
\subsection{Key Concepts}

\begin{itemize}
    \item \textbf{Borg}: a system for cluster management at Google. See \citep{verma2015large} for more information. Borg monitoring systems provide resource usage and power consumption data, which we use to estimate machine energy and allocate it to users.
    \item \textbf{Busy power}: the maximum power draw from a machine while running actual workloads in a data center.
    \item \textbf{Cluster}: a group of production machines that (typically) share a common local network infrastructure and are located in a single datacenter building. 
    \item \textbf{Dedicated machine:} A datacenter machine with a single machine owner, a single production group that controls access to and operates the machine. Non-dedicated machines are called shared machines.
    \item \textbf{Dynamic power}: the measured power draw from a machine, minus the machine's idle power.
    \item \textbf{GCU}: Google Compute Unit, a platform neutral measurement for central processing unit (CPU) resources \citep{dev2020autonomous}. Different processors are assigned a GCU-per-core rating by bench-marking the processor's performance. In the pipeline implementation, when CPUs of different machines have different performance, we use the GCU concept to normalize across different CPUs.
    \item \textbf{HDD}: Hard disk drive, responsible for data storage. 
    \item \textbf{Idle power}: the minimum power draw from a machine when it is active in a data center, for example if it is powered on but not running any workloads.
    \item \textbf{Location-based emissions}: Carbon footprint based on electricity consumption in Google's datacenters, \emph{not} accounting for any direct clean energy purchases made by Google.
    \item \textbf{Machine owner}: for dedicated machines, the single production group that controls access to the machine.
    \item \textbf{Market-based emissions}: Carbon footprint based on electricity consumption in Google's datacenters, accounting for clean energy purchases made by Google.
    \item \textbf{Production group:} Permissions groups used for production services. These are groups that control resource usage.
    \item \textbf{Production job}: High-priority machine tasks, for example, long-running server jobs, which take priority in Borg over lower-priority non-production jobs (often batch jobs). 
    \item \textbf{PUE}: total datacenter power divided by information technology (IT) power.
    \item \textbf{RAM}: Random-access memory.
    \item \textbf{Resource allocation}: The amount of a particular resource that is reserved for a particular production group or user. Production jobs will often have resource allocations higher than resource usage, so they can easily scale up (pushing out non-production jobs) to meet external demands. See Section 5.5 in \citep{verma2015large} for more information.
    \item \textbf{Resource usage}: The amount of a particular resource, e.g. GCUs, GiBs of memory, or TiBs of HDD, used by a particular production group or user. See Section 5.5 in \citep{verma2015large} for more information.
    \item \textbf{Shared machine:} A datacenter machine with a multiple internal users.
    \item \textbf{Shared service}: An internal product whereby a team provides a combination of software and resources (i.e. disk space and the tools to easily store data in a redundant manner) to other internal users. Examples include Bigtable and Spanner.
    \item \textbf{SSD}: Drive that uses solid-state memory to store persistent data.
    \item \textbf{User}: A production group or collection of production groups. For example, the user Compute Engine refers to a group of production groups that build and operate the public Google Compute Engine (GCE) product. Similarly, you can think of Google Search or YouTube, or a sub-components of those teams, as being a single user.
\end{itemize}

\subsection{Boundaries and Exclusions}
Google's data center energy and emissions measurement is the approach that forms the backbone of the Scope 2 Google Cloud Carbon Footprint reporting methodology, and as such, the boundaries and exclusions are the same as are described in our public methodology \citep{GCPmethodology}.\footnote{The Google Cloud Carbon Report also includes estimates of per-customer emissions related to Google's Scope 1 and Scope 3 emissions. Those estimates are described in \citep{GCPmethodology}. This paper focuses on the Scope 2 methodology so the boundary of this paper only includes activities related to Scope 2 emissions.}

The inventory boundary we use is for Scope 2 emissions resulting from electricity use by Google-owned or Google-operated data centers, including that from compute and networking equipment as well as ancillary electricity services such as cooling and lighting.

The Scope 2 Carbon Footprint report excludes emissions arising from the following activities:
\begin{itemize}
    \item Generating electricity that is subsequently lost during transmission and distribution.
    \item Extraction and transportation of fuels used to generate grid electricity, and the life-cycle emissions associated with the generation facilities and equipment.
    \item All Scope 1 emissions, including for example fugitive emissions from HVAC system coolants.
    \item Emissions arising from small equipment deployments at internet service providers' partners.
    \item Emissions from Google networking equipment deployed outside data centers.
    \item Downstream end-of-life emissions of data center equipment and buildings.
\end{itemize}

\subsection{Machine Power Allocation} \label{sec: power allocation}

Google's data centers support a large variety of machine types and workload types. A few features of warehouse-scale datacenters make an allocation of machine power to users especially complicated:

\begin{enumerate}
    \item Google datacenter clusters include \emph{dedicated machines}, which are operated by a single user, and \emph{shared machines}, which can be operated by multiple users, even simultaneously. 
    \item In a datacenter cluster, resources are shared by production and non-production jobs. At Google, production jobs are allocated about 70\% of the total GCU resources and represent about 60\% of the total GCU usage; they are allocated about 55\% of the total memory and represent about 85\% of the total memory usage \citep{verma2015large}.
    \item Workloads are highly heterogeneous, with different usage patterns and resource requirements. While higher utilization of machines can help reduce costs and energy consumption, datacenters may require additional machines to meet daily and seasonal demand patterns or sudden increases in resource demand. 
    \item Server energy consumption is not \emph{proportional}. While idle, a server on average uses approximately 45\% of its peak power consumption, or approximately 60\% of its average energy consumption \citep{fan2007power}. If one task is running on a machine at 40\% utilization and it is joined by a second task with the same resource requirements, doubling utilization to 80\%, power consumption will not double; rather it will go up by approximately 30\%. 
\end{enumerate}

To balance these complexities with the need for a simple and explainable approach, we separate the allocation of idle power and dynamic power for shared machines. Reported idle power is approximately constant for each machine--it is the power a machine consumes even if it's not in use. However, at the data center level, idle power can vary hour-to-hour based on various changes like machine failures, upgrades, and down-time. Dynamic power is the additional power consumed, above and beyond idle power, and it can vary significantly as the utilization or job requirements on a particular machine can change (see, e.g Figure \ref{fig:cluster_ia_power} below). Let $p_{m,h}$ be the measured power and $i_{m,h}^r$ be the recorded idle power of machine $m$ in hour $h$, respectively. Measured power is sampled every five minutes, with the hourly measured power $p_{m,h}$ being the average sampled value of the five minute intervals. We define and calculate idle power as follows:
\begin{equation}\label{idle_eq}
    i_{m,h} = \min \{ i_{m,h}^r , p_{m,h} \}.
\end{equation}
Then, dynamic power is calculated as:
\begin{equation}\label{dynamic_eq}
    d_{m,h} = p_{m,h} - i_{m,h}.
\end{equation}
The adjustment accounts for potential mis-configured idle power readings, ensuring that $i_{m,h} \leq p_{m,h}$ and $d_{m,h} \geq 0$. 

Next, we allocate the measured power to internal Google users. The overall approach is as follows:
\begin{enumerate}
    \item Allocate idle power of dedicated machines to the machine owner.
    \item Allocate idle power of shared machines proportional to a power-weighted blend of resource allocations (of GCU, HDD, SSD, RAM) in the cluster where the shared machine is located.
    \item Allocate dynamic power of a machine to the real-time machine users, proportional to their GCU usage on the machine in the particular hour under consideration.
\end{enumerate}

The basic idea of our approach is that demand for resource allocations--of dedicated machines, GCUs, RAM, HDD, and SSD--drives idle power of machines, since demand for resource availability drives the deployment of machines. The actual usage of these machines drives their dynamic power; if machines are not used, they consume power at a rate approximately equal to their idle power rating.

This approach helps address the four main issues outlined above. First, it enables different treatment of dedicated and shared machines, and it allows for the power of shared machines to be allocated to and divided across multiple users. Second, it suggests a reasonably fair way of balancing the impact of production and non-production jobs, where production jobs have higher footprint due to their associated resource allocations. Third, it handles heterogeneous jobs and resource requirements by having idle power distributed proportional to a blend of resource-types. Finally, even though server energy consumption is not proportional to CPU usage, dynamic power is more closely proportional to CPU usage, so by separating the allocation mechanisms for idle and dynamic power we can more safely assume that dynamic power is roughly proportional to resource usage.\footnote{\revb{Note that we are not assuming that dynamic power is proportional to CPU usage. Rather, we measure dynamic power directly and argue that the dynamic power can be fairly allocated to users proportional to their CPU usage, when there are multiple users.}}

\begin{figure}[h]
\centering
\includegraphics[scale=0.60]{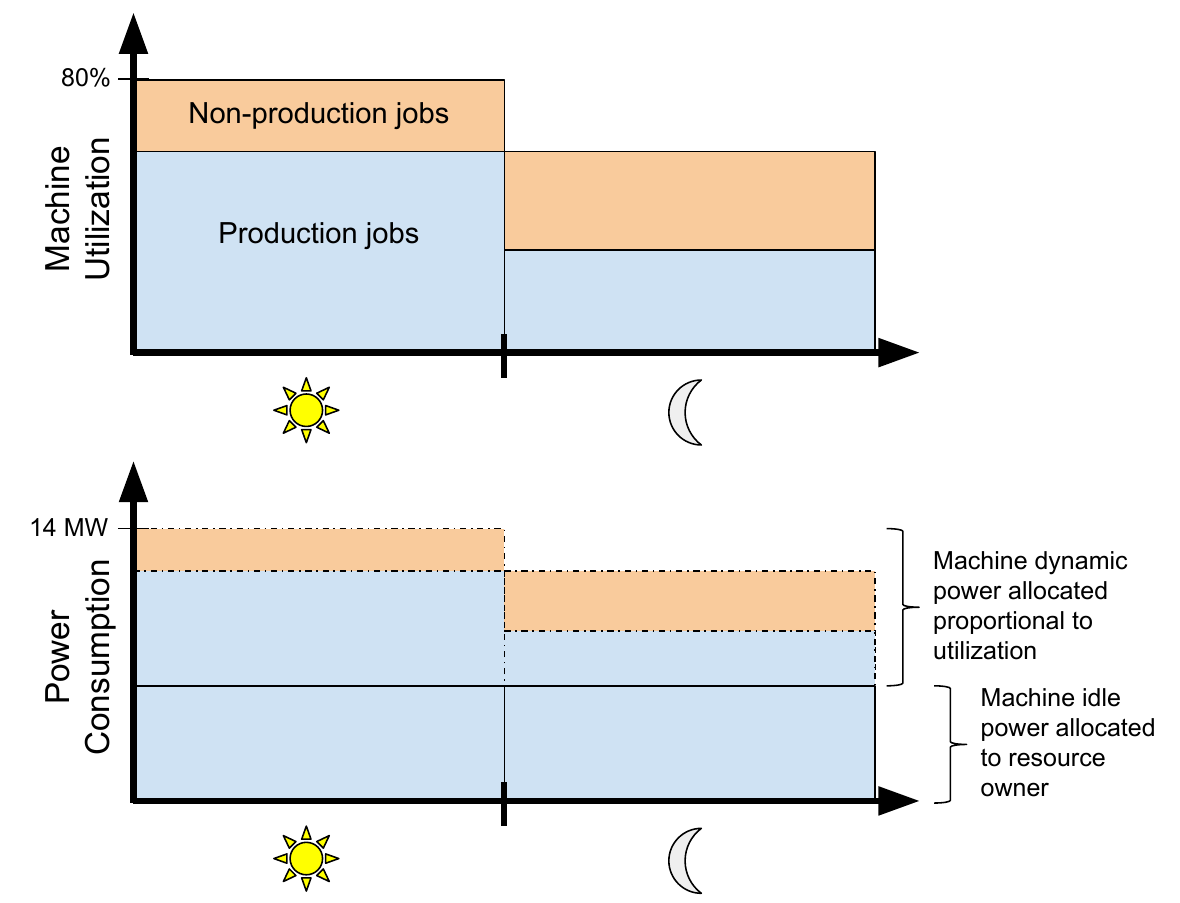}
\caption{A simplified example of energy allocation in a datacenter with diurnal usage patterns and no shared services. One user (blue) is responsible for all production jobs and resource allocations; the other (orange) runs non-production jobs. In the upper chart, the example utilization values would be measured. In the lower chart, power is measured, while the proportion of power to allocate to each user is calculated based on utilization measurements and the described formulas.}
\label{fig:simple_example}
\end{figure}

Before we detail the approach, consider a simplified example detailed in Figure \ref{fig:simple_example} to highlight the main features: a 16MW data center cluster with a single user running production jobs and a single user running non-production jobs. 
Assume we measure that utilization of the machines is 80\% during the day and 60\% at night, with the prod user contributing 75\% of of the utilization during the day and 50\% of the utilization at night. Assume that the prod user has all of the resource allocation for this cluster: they have control of the resources and can run when they need to, usurping the non-prod user. Assume also that the idle power of the machines is 6MW, but the measured power consumption from the machines is 12 MW at 60\% utilization and 14MW at 80\% utilization. Then, under our allocation rules, the prod user is assigned the 6MW idle power during all hours of the day. The two users split the dynamic power proportional to their utilization. In all, the prod user is allocated 12 MW during the day and 9 MW at night. The non-prod user is allocated 2 MW during the day, 3 MW at night. 

Figure \ref{fig:cluster_ia_power} is a graph of idle, dynamic, and total power for a single cluster of servers from September 18 - September 24, 2023.  Note that the amount of idle power is (slightly) larger than dynamic power, and remember that total power is the sum of idle power and dynamic power.

\begin{figure}[h]
\centering
\includegraphics[scale=0.2]{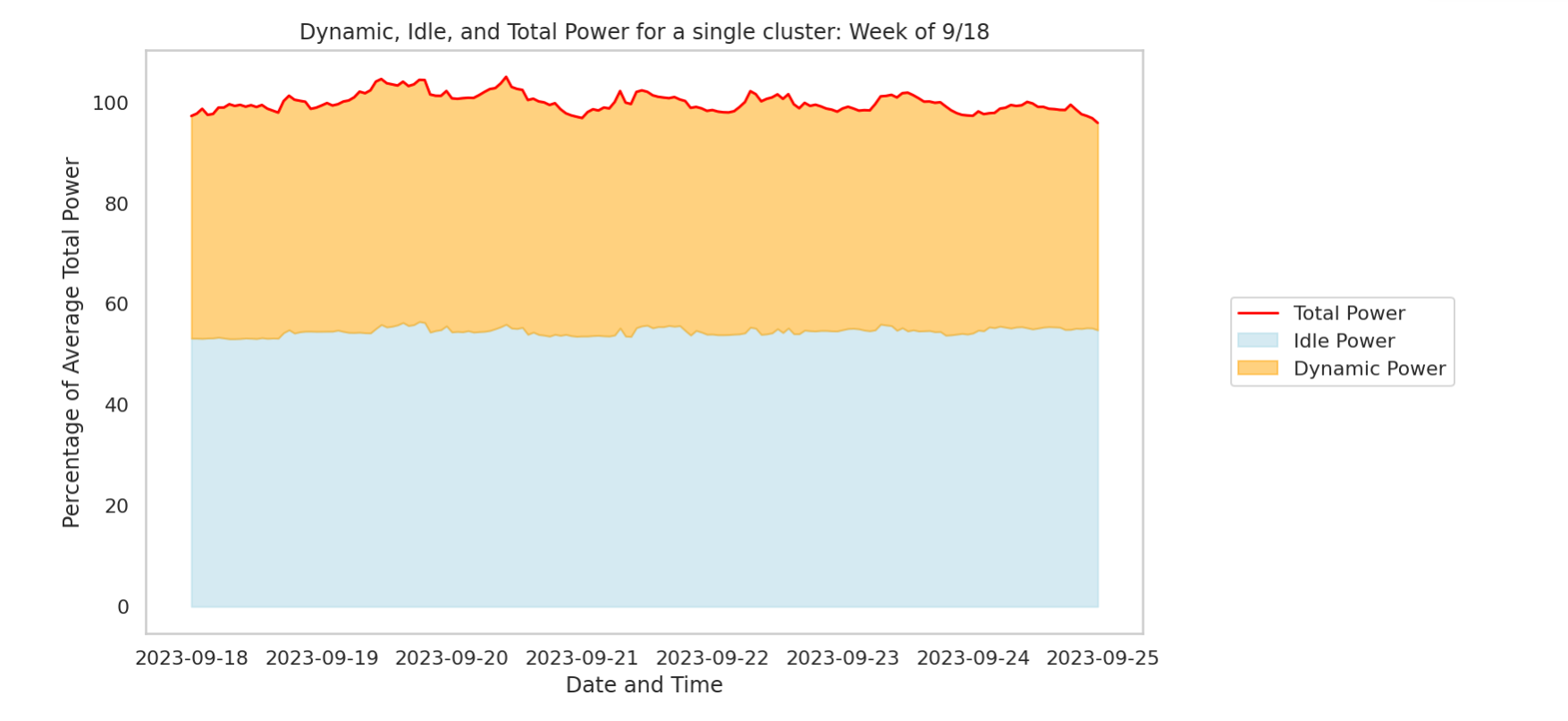}
\caption{The amount of idle power, dynamic power, and total power consumed per hour for a specific Google Cloud cluster, as a percentage of the weekly average total power.}
\label{fig:cluster_ia_power}
\end{figure}

\begin{figure}
\centering
{\includegraphics[scale=0.17]{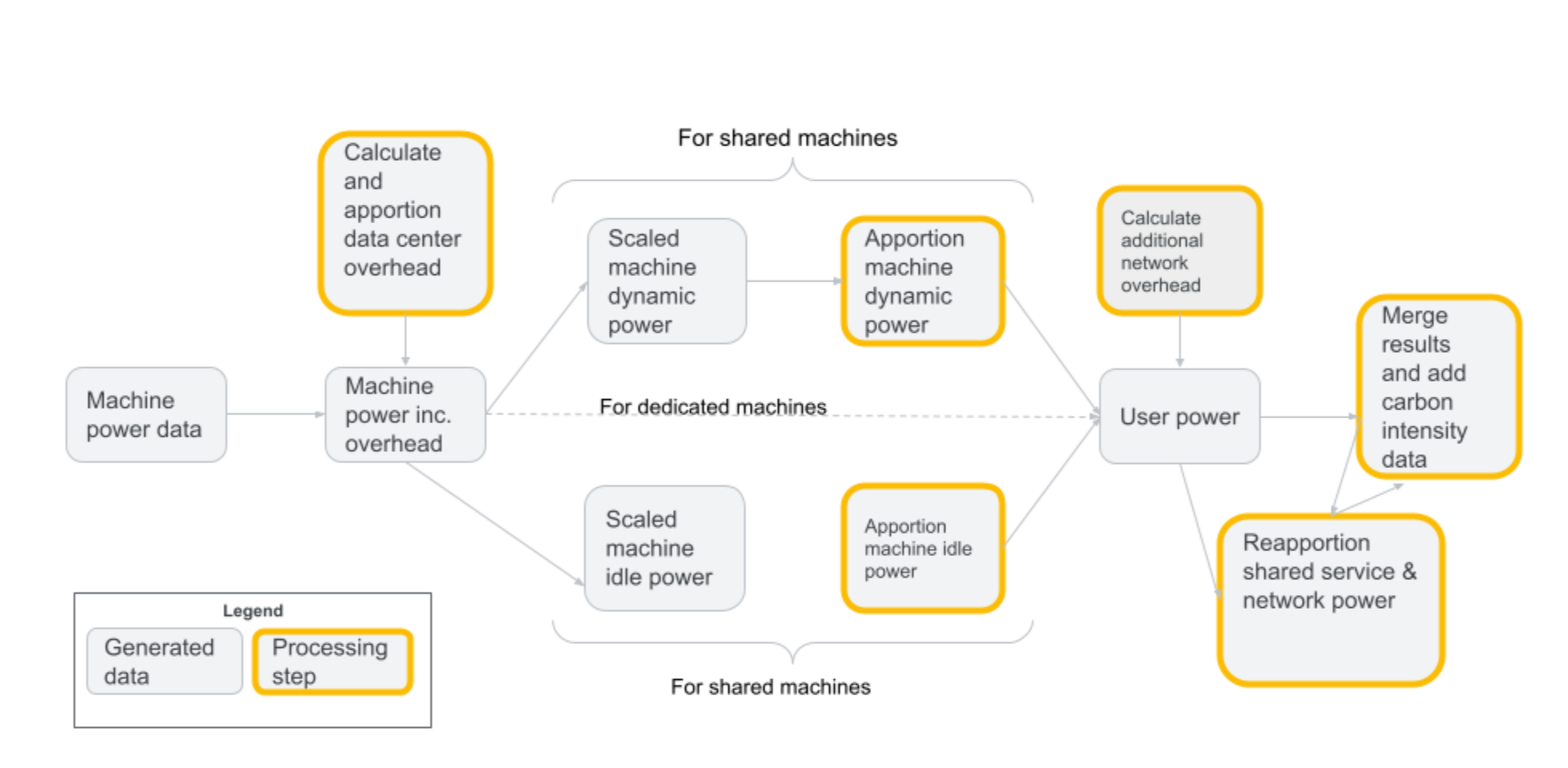}}
\caption{High-level overview of the process of allocating energy and carbon emissions to internal Google users.}
\label{fig:pipeline_design}
\end{figure}

The following sections detail the full methodology used to allocate energy and emissions to internal users. Figure \ref{fig:pipeline_design} provides a detailed flow chart of the overall process that is described in detail in those sections.

\subsubsection{Machine Idle Power Allocation} \label{sec: idle_allocation}

For both dedicated and shared machines, idle power is measured and calculated according to equation \eqref{idle_eq}. Previous studies have shown that idle power corresponds to approximately 60\% of average server energy consumption \citep{fan2007power}, with dynamic power (as defined by \eqref{idle_eq}) corresponding to the remaining 40\% of server energy consumption.

We use a different methodology to allocate that idle power, depending on whether the machine is a dedicated or shared machine. For dedicated machines, we allocate the idle power of the machine to the machine owner. The machine owner is a specific production group that is labelled as the owner of the machine in Google's internal data. Dedicated ownership of machines is common for users that require isolation or more flexibility regarding the machine types they deploy; example users are ML Infrastructure and Google Cloud's Compute Engine. For dedicated machines, power allocation is simple. We look at an internal tag which denotes machine owner--the production group that owns these machines--and assign all of the idle power of these machines to that production group. This does not imply that all of the energy consumption and carbon footprint associated with these machines will be associated with the machine owner in our final data set, because if the machine owner operates an internal shared service, some of that energy and carbon may be reallocated (see Section \ref{sec: reallocating energy}) to the users of that internal shared service.

For shared machines, we allocate idle power of all machines in each cluster proportional to the power-weighted resource allocation of each user in the cluster. Let $S$ be the set of shared machines. For this section, consider a particular hour $h$; in each equation, we drop the subscript $h$ for simplicity. Then, for a particular cluster $C$ the total idle power of the shared machines is  $\sum_{m \in C, m \in S} i_m$. For a particular user $j$, their fraction of the cluster idle power, $f_{j,C}$ is given by the following equation:
$$ f_{j,C} = \frac{c^{GCU} r^{GCU}_{j,C} + c^{RAM} r^{RAM}_{j,C} + c^{HDD} r^{HDD}_{j,C} + c^{SSD} r^{SSD}_{j,C}}{\sum_{j} ( c^{GCU} r^{GCU}_{j,C} + c^{RAM} r^{RAM}_{j,C} + c^{HDD} r^{HDD}_{j,C} + c^{SSD} r^{SSD}_{j,C})}$$
For some resource type $R$, $c^{R}$ is the power weighting of the resource and  $r^{R}_{j,C}$ is user $j$'s allocation of shared resource $R$ in cluster $C$. User $j$'s allocation of shared machine idle power in cluster $C$ is the product of two terms, $f_{j,C}$ and the total idle power of the shared machines, $\sum_{m \in C, m \in S} i_m$. Note that the equation does not include allocation of any ML-specific resources, because, as mentioned above, nearly all ML machines are operated as dedicated machines. 

The power weightings come from internal estimates of the busy power associated with each resource type. While it might be more accurate to use internal estimates of idle power for this use-case, we use busy power because its readily available internally and because it has multiple internal use-cases so we have higher confidence that the data can be maintained as accurate and current. In either case, the goal is to get a power-weighting to normalize and combine allocations across different metric types. The following chart summarizes the weightings by showing approximately how many units of each resource draws the equivalent power of 1 CPU core; we refresh the internal data approximately quarterly, and new machine generations often lead to marked improvements, as evidenced by the previously mentioned fact that, as of 2022, Google delivered approximately three times as much computing power with the same amount of electrical power as it did five years prior \citep{googleER}.

\begin{table}[H]
\centering
\begin{tabular}{|ll|}
\hline
\multicolumn{2}{|l|}{Resources with equivalent busy power} \\ \hline
\multicolumn{1}{|l|}{1}                    & Intel Skylake N1 CPU Core                       \\ \hline
\multicolumn{1}{|l|}{20}                 & GiB RAM                    \\ \hline
\multicolumn{1}{|l|}{1}                & TiB SSD                    \\ \hline
\multicolumn{1}{|l|}{6}                & TiB HDD                    \\ \hline
\end{tabular}
\label{tab:my-table}
\end{table}


In summary, the total idle power allocated to user $j$ in cluster $C$ is
$$I_{j,C} = \sum_{m \in C, m \in D_j} i_m + f_{j,C} \sum_{m \in C, m \in S} i_m.$$
The first term is the sum of idle power of the dedicated machines owned by user $j$ in cluster $C$, where $D_j$ is the set of dedicated machines owned by user $j$. The second term is user $j$'s allocation of idle power from shared machines in cluster $C$ as explained above.

\subsubsection{Machine Dynamic Power Allocation} \label{sec: dynamic_allocation}

For each machine, we calculate the dynamic power in each hour, using equations \eqref{idle_eq} and \eqref{dynamic_eq}. We allocate dynamic power to users proportional to their fraction of the machine's host GCU usage within that hour window. Let $GCU_{i,m}$ be the GCU usage by user $i$ on machine $m$ (in a particular hour $h$, but dropping the $h$ for simplicity). Then the dynamic power for a particular user $j$ in cluster $C$ is  

$$D_{j,C} = \sum_{m \in C} \left( d_m \frac{GCU_{j,m}}{\sum_i GCU_{i,m}}\right).$$

User $j$'s dynamic power allocation in cluster $C$ is the sum of their dynamic power allocation for each machine in cluster $C$ for which they have non-zero GCU usage. On each machine, dynamic power is allocated to users proportional to their GCU usage in that particular hour. 

The approach, allocating dynamic power proportional to GCU-usage, was chosen for its relatively accuracy and simplicity, but it does not capture all of the nuances of machine dynamic power usage. For example, if two users each use half of the GCU capacity on a machine, but the second user has much higher memory usage, then in reality the second user is probably responsible for more than half of the machine's dynamic power usage. In our view, this means that we are possibly underestimating the true dynamic power allocation for users with above-average memory usage. 

These shortcomings could be especially evident on machines with significant non-CPU resources, like storage machines and ML machines. Fortunately, the typical usage patterns and physical characteristics of each of those machine types mitigate potential shortcomings. While storage is occasionally directly attached to compute servers, most storage disks are on separate network-attached storage machines \citep{barroso2019datacenter}. These storage machines typically have a low dynamic range, so most of their power is allocated as idle power. Most ML machines are dedicated machines and frequently all of their dynamic power consumption is allocated to a single user.

\subsection{Reallocating Energy of Internal Services} \label{sec: reallocating energy}

At Google, many teams build services that other teams use and rely on. For example, Colossus \citep{colossus_description} \citep{colossus_description}, is the storage system that teams throughout Google rely on to use and access network-attached storage for their various products. For carbon footprint accounting, we strive to reallocate the energy and carbon from these services to their end-users at Google. For example, if Gmail uses Colossus, our default view of the data is one in which Gmail's fraction of Colossus energy usage is reallocated from Colossus to Gmail. We also can provide a view on the data without the reallocation, which can be helpful for service providers (like the team that builds Colossus) to understand their carbon emissions, but for most users and for external reporting we reallocate usage to the end-users. 

We perform the reallocation of internal service energy based into two distinct approaches, depending on the data available. For a few of the largest services, we have detailed data on the resource usage that their users require. The reallocation of those services is described in \ref{padm_reallocation}. For the long-tail of smaller services, we have to rely on internal cost data that's used internally for budgeting and accounting. The reallocation of these services is described in Section \ref{net_cost_reallocation}. 

\subsubsection{Reallocating Energy of Major Shared Services} \label{padm_reallocation}
For major internal shared services, like Colossus, Spanner, and Bigtable, we have direct information about the resource allocation and resources used on the service by the users of those services. We use this data to reallocate both idle power and dynamic power from major shared services to users. 

For idle power, the approach is simple, because resource allocation at Google is already attributed to users of these major shared services. For large services, internal resource allocation modelling includes a mapping from shared services to the underlying base resources used to provide them, so our internal data has resources allocated directly to the end user instead of the shared service.  For example, 1 Bigtable server and 20 TiB translates into $s$ GCUs, $r$ RAM and $t$ HDD, so a user consuming 1 Bigtable server and 20 TiB has the additional resource allocations for $s$ GCUs, $r$ RAM and $t$ HDD in our internal data source. Thus, the resource allocation value $r_{j,C}^R$, for each resource $R$, cluster $C$, and user $j$ described in Section \ref{sec: idle_allocation} already includes the reallocation of resource allocations from shared services to users. For example, if Gmail uses Colossus, any resource allocations of HDD associated with Gmail's use of Colossus in any cluster $C$ are already included in $r_{Gmail, C}^{HDD}$. The corresponding idle power is already attributed to Gmail using the approach in Section \ref{sec: idle_allocation}.

On the other hand, dynamic power consumption must be reallocated from the shared service to its users because the approach in Section \ref{sec: dynamic_allocation} relies on data about task usage which is associated with the shared service team that is running the compute task. 

Similar to $r_{j,C}^R$, let $u_{j,k,C}^R$ be user $j$'s usage of resource $R$ for a shared service managed by a different user $k$ in cluster $C$. We reallocate the following fraction of dynamic power from user $k$ to user $j$ in cluster $C$:

\begin{equation} \label{dyn_reallocation}
    \gamma_{j,k,C} = \frac{u^{GCU}_{j,k,C}}{\sum_{l} u^{GCU}_{k,l,C}}.
\end{equation}

The numerator is user $j$'s fraction of GCU usage from the service owned and operated by user $k$. The denominator is user $k$'s total GCU usage across all products and services. 

For Colossus \citep{colossus_description, serenyi2017cluster}, a more complicated approach is required because as a storage system, much of the resource usage associated with Colossus comes from hard-disk resources. Thus, for Colussus, we use the usage-power weighted-resource fraction of GCU, HDD, and SSD. The usage-power associated with each resource $R$ is defined as $\Tilde{c}^{R}$; it is the average power required for one additional unit of usage of the resource. This is similar to the power weighting described above $c^R$, except it is the incremental average power consumption associated with one unit of resource usage, instead of the incremental busy power associated with one unit of resource allocation. 

We allocate the following fraction of dynamic power from Colossus, denoted here as user $w$, to each user $j$ in each cluster $C$:

\begin{equation} \label{dyn_reallocation_colossus}
    \gamma_{j,w,C} = \frac{\Tilde{c}^{HDD} u^{HDD}_{j,w,C} + \Tilde{c}^{SSD} u^{SSD}_{j,w,C} + \Tilde{c}^{GCU} u^{GCU}_{j,w,C}}{\sum_{l} \Tilde{c}^{HDD} u^{HDD}_{w,l,C} + \Tilde{c}^{SSD} u^{SSD}_{w,l,C} + \Tilde{c}^{GCU} u^{GCU}_{w,l,C}}.
\end{equation}

\subsubsection{Reallocating Energy of Minor Shared Services} \label{net_cost_reallocation}
Smaller shared services cannot be reallocated using the methodology described in Section \ref{padm_reallocation}, because we do not have granular data on which resources their users are consuming when they utilize the shared service. Unlike major shared services, we do not have direct information about how many resources were consumed by users of these smaller shared services. The methodology described in the previous Section \ref{padm_reallocation}, is used for approximately 2/3 of shared service energy consumption, while the methodology in this section is used for approximately the remaining 1/3 of shared service energy consumption. 

For these smaller shared services, we reallocate power and carbon globally to users/production groups, proportional to a \textit{net cost} metric. The \textit{net cost} metrics are derived from Google's Resource Economy, which is a set of teams, policies, and processes that Google uses to plan and deliver our compute, storage, and network infrastructure resources. They represent the cost of resources a user prompted Google to acquire. These are \textit{derived} metrics, based on consumption on the consumer side, and capacity data on the inventory side. The data is used to enable teams at Google to make financial decisions about their own products by exposing economic data about the cost of shared services, compute resources, and networking.

 For our purposes, we have data on the \textit{net cost} per production group,  shared service, and day. For production groups that use the shared service, net cost is positive, whereas for production groups that provide the shared service, net cost is negative. We also have idle and dynamic power per service and per production group as outlined in Section \ref{sec: power allocation}. To reallocate power and carbon for these services, we first identify the primary provider (a specific production group or set of production groups) for each shared service. Next, we calculate the fraction of the shared service provider's power and carbon that should be reallocated to each of their service's users. This accounts for the fact that a few providers, like Google Cloud's Compute Engine, provide multiple services. For each production group, we re-attribute power and carbon proportional to their net cost of the service. For example, if Google Ads pays \$1,000 to Blobstore, a system for object storage at Google \citep{google_object_storage}, which has total revenue of \$10,000 on a particular day, then Ads is allocated 10\% of Blobstore's footprint at every location on that day. 

Let $n^S_j$ be the net-cost to user $j$ from shared service $S$. This number can be negative if the user $j$ receives revenue from service $S$. In particular, we say that user $k$ is the provider of service $S$ if
\begin{equation}\label{service owner eq}
    k = \arg\min_i n^S_i.
\end{equation}
\reva{User $k$ is the provider of service $S$ if they are the user that receives the most revenue from Service $S$. Typically, only one user receives revenue from any service, so for any service $S$ and any other user $i$, $n^S_i \geq 0$ and for a single service provider $k$, $n^S_k < 0$.} Furthermore, let $\hat{n}_k$ be user $k$'s non-shared-service costs. Then, the total costs for user $k$ are $TC_k = \hat{n}_k + \sum_{S} n^S_k$. Now, let $\overline{TC}_k = \max \{|n^S_k|, TC_k \}$. We reallocate $n^S_j / \overline{TC}_k$ of user $k$'s energy consumption to user $j$ of service $S$.\footnote{We use $\overline{TC}_k$ in the denominator instead of $TC_k$ to handle edge cases in the internal accounting whereby a user might have higher internal revenues than costs. This change ensures that at most 100\% of the power originally allocated to user $k$ is reallocated to its service's users.}

This approach has a few desirable properties. First, power is reallocated to users proportional to their costs associated with each specific service. In carbon accounting, a cost-based approach is typically less desirable than a resource-based approach, but we are able to mitigate some of the downsides of the cost-based approach by applying it to each specific shared service for higher granularity. Second, in the ``balanced'' accounting case where a service is used entirely internally, all of the energy consumption originally allocated to the service is reallocated to other internal users. In this case, the services costs equal its internal revenues, so $|n^S_k| = TC_k$ and thus $\overline{TC}_k = |n^S_k|$. The accounting cost-balance requires that $n^S_k = - \sum_{j \neq k} n^S_j$, and  $n^S_k < 0$, so in this case the total fraction of energy reallocated to users is $\sum_{j \neq k} n^S_j / \overline{TC}_k = - n^S_k / |n^S_k| = 1$. Third, in the case where a service has costs greater than revenue, a fraction of its power is reallocated to users and some is retained by the shared service provider $k$, proportional to the relative fraction of its costs and internal revenues. 

\begin{figure}[h]
\centering
\includegraphics[scale=0.80]{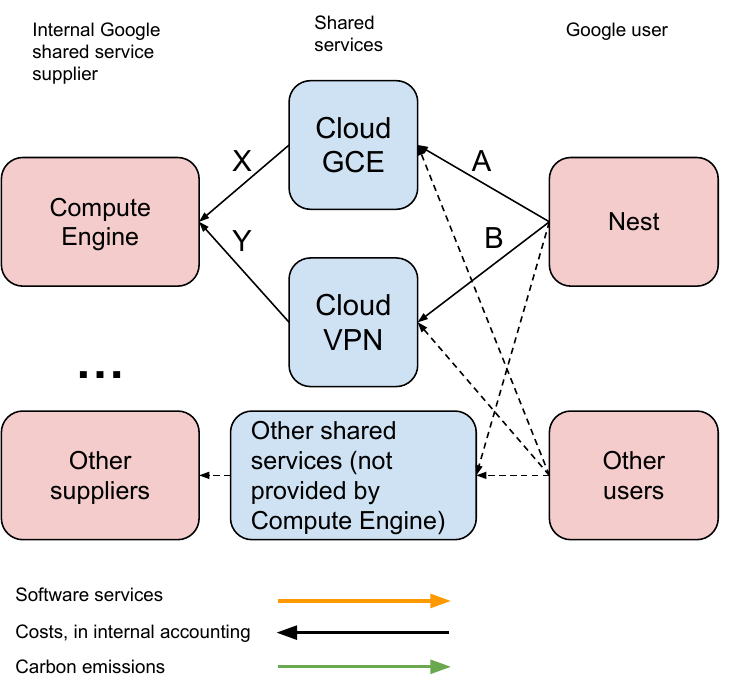}
\caption{Example of net cost flows in the resource economy used to allocate carbon of internal Google shared services.}
\label{fig:net_cost}
\end{figure}

For a working example, please see below. Blue boxes represent shared services, red boxes represent users, and black lines represent dollar flows. Software services (not depicted) are provided by users on the left to users on the right. Accounting costs flow from users on the right to users on the left. Carbon is reallocated, based on those accounting costs, to users on the right from users on the left. A single example user is depicted: Nest, which builds smart-home hardware and software. Nest is charged (on an accounting-basis)  $A = n^{GCE}_{Nest}$  for its use of the shared service Cloud GCE and $B = n^{VPN}_{Nest}$ for its use of the shared service Cloud VPN. Compute Engine (itself an internal Google user, within Google Cloud) manages both of these shared services. Of all of the revenue that Compute Engine receives from various shared services, its net cost in total is $TC_{Compute Engine}$, which is its total accounting costs for machines, other resources, and other shared services, minus its revenues for the services it provides, e.g. $X + Y$ in Figure \ref{fig:net_cost}. Then, in each cluster where Compute Engine has a footprint, we reallocate $A / TC_{Compute Engine}$ of Compute Engine's footprint to Nest for its use of Cloud GCE, and we reallocate $B / TC_{Compute Engine}$ of that Compute Engine's footprint to Nest for its use of Cloud VPN. 

The diagram names only one user, Nest. In practice, Nest would consume other shared services provided by Compute Engine in addition to Cloud GCE and and Cloud VPN and shared services not supplied by Compute Engine. Additionally, many other users (depicted at bottom) would use the same set of shared services both supplied by Compute Engine and supplied by other internal users,  but the same logic applies.

We run this allocation \emph{twice} to cover situations where Shared Services use other shared services. This pair of computations occurs after the first round of reallocation described in Section \ref{padm_reallocation}, so there are three total rounds of Shared Service reallocation. For example, Blobstore is allocated carbon from Colossus (in the first round of reallocations described in \ref{padm_reallocation})). Then, this energy is reallocated using net cost to Cloud Storage (in the first round of cost-based reallocation described here). Then, the energy/carbon that Cloud Storage is assigned from Blobstore is reallocated to Cloud Storage's internal users during the second round of reallocation. \revb{Additional rounds of allocation would lead to a more accurate final result, at the cost of additional query resources because the query joins required to allocate energy for the entire fleet are very large. We implemented the pipelines with two rounds of allocation because in testing we observed little change in emissions beyond two rounds of allocation. Additional rounds of allocation would lead to slightly higher accuracy in allocation but would require additional pipeline resources and processing time.}

\subsubsection{Putting it All Together}
To get the overall allocated energy consumption per user, we run the following algorithm to calculate the energy footprint of each user $j$ in each cluster $C$.

\begin{enumerate}
    \item For each user $j$ and cluster $C$, calculate idle power $I_{j,C}$. As mentioned above, this naturally encapsulates the reallocation of shared service idle power to users because that is done in our upstream data source for resource allocations.
    \item For each user $j$ and cluster $C$, calculate dynamic power $D_{j,C}$.
    \item For each user $j$ and each user $k$ that operates a major shared service in each cluster $C$, calculate $\gamma_{j,k,C}$ using \eqref{dyn_reallocation} and \eqref{dyn_reallocation_colossus} (the latter, just for Colossus, and the former for the remaining services). For each pair and cluster, reallocate $\gamma_{j,k,C} D_{k,C}$ from user $k$ to user $j$. Let $P^A_{j,C}$ be the new total power consumption for user $j$ in cluster $C$, comprising idle power, dynamic power, and reallocated dynamic power.
    \item For each user $j$ and each user $k$ that operates a minor shared service, reallocate $\left( n^S_j / \overline{TC}_k \right) P^A_{k,C}$ from user $k$ to user $j$. For any user $j$ let $P^B_{j,C}$ be the new total power consumption after the reallocation. 
    \item Finally, repeat the previous step. For each user $j$ and each user $k$ that operates a minor shared service, reallocate $\left( n^S_j / \overline{TC}_k \right) P^B_{k,C}$ from user $k$ to user $j$. For any user $j$, let $P_{j,C}$ be the final total power consumption after the reallocation. 
\end{enumerate}

The final result is the energy consumption $P_{j,C}$ of each user $j$ in every cluster $C$, accounting for their machine usage, resource allocation, and usage of dozens of shared services within Google's global compute ecosystem.

A high-level overview of the end-to-end process was shown above in Figure \ref{fig:pipeline_design}. In addition, the following Figure \ref{fig:sankey} highlights the energy allocations for Google Cloud from the overall process. The first column shows the relative breakdown of machine power and overhead datacenter energy consumption. The overhead (covered in the next section, \ref{sec: pue and carbon}) is applied proportionally to both machine idle and dynamic power, shown in the second column. The third column shows how energy is allocated, from top-to-bottom, based on real-time machine utilization (Section \ref{sec: dynamic_allocation}), dedicated machine ownership (Section \ref{sec: idle_allocation}, and allocation of shared machine resources (Section \ref{sec: idle_allocation}. The fourth column shows the reallocation of energy from major and minor shared services \ref{sec: reallocating energy}. For simplicity, we're only including energy flows that impact Google Cloud and combining all other internal Google users that provide shared software services to Cloud or use Google Cloud's internal software services. In reality, this diagram only shows a fraction of Google's overall data center usage.

\begin{figure}[h]
\centering
\includegraphics[scale=0.41]{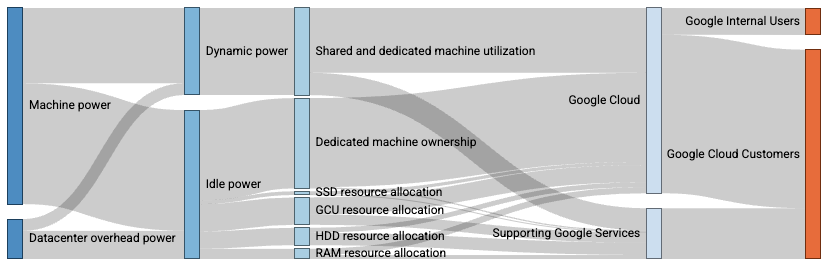}
\caption{Sankey diagram showing illustrative energy flows allocated to Cloud and internal Google users of Google Cloud. This represents, using an illustrative example, a portion of the data center energy consumption at Google that pertains directly to Google Cloud.}
\label{fig:sankey}
\end{figure}

\subsection{Incorporating energy overheads and carbon intensity} \label{sec: pue and carbon}
Now that we have hourly energy consumption allocations to each user in each location, we can combine that information with data center energy overheads and grid carbon emissions intensity values to estimate the ``location-based'' carbon footprint of each user in each of Google's datacenters, which does not account for any clean energy purchases made by Google.

Energy overhead refers to the additional energy required to run non-IT equipment in datacenter. For example, this includes energy required for cooling, lighting, ancillary office equipment, and some power losses.  Power Utilization Effectiveness (PUE) is a metric which captures total power / IT power for a particular data center. Google has internal information on hourly PUE for its own datacenter and estimated PUE for third-party data centers we use. 

Carbon intensity is a measure of carbon emissions per unit of electricity production or consumption. Our methodology uses operational, not life cycle assessment (LCA) values. Carbon free energy sources such as wind, hydro, or solar power, produce little if any operational CO$_{2}$ emissions. Therefore, the operational carbon intensity of electricity that is generated from carbon-free energy sources is approximately 0 gCO$_{2}$e/kWh.

To convert the energy consumed by Google's internal services into an estimate of carbon emissions, we utilize data from Electricity Maps. They provide carbon intensity data associated with electricity consumption for about 200 zones per hour \citep{tranberg2019real}. A zone is defined as the physical grid on which consumers and producers are connected and which is controlled by a grid operator.\footnote{https://github.com/electricitymaps/electricitymaps-contrib/wiki/What-is-a-zone} Electricity Maps aggregates data from government and government affiliated sources (energy ministries, official statistical bureau, etc), transmission or Distribution System Operators, and utility companies that generate or manage power directly. The carbon intensity metrics that Google uses measure the carbon intensity that arises from electricity production (Scope 2), and not other life cycle stages of the plant.

\begin{figure}[htbp]
\centering
\includegraphics[scale=0.25]{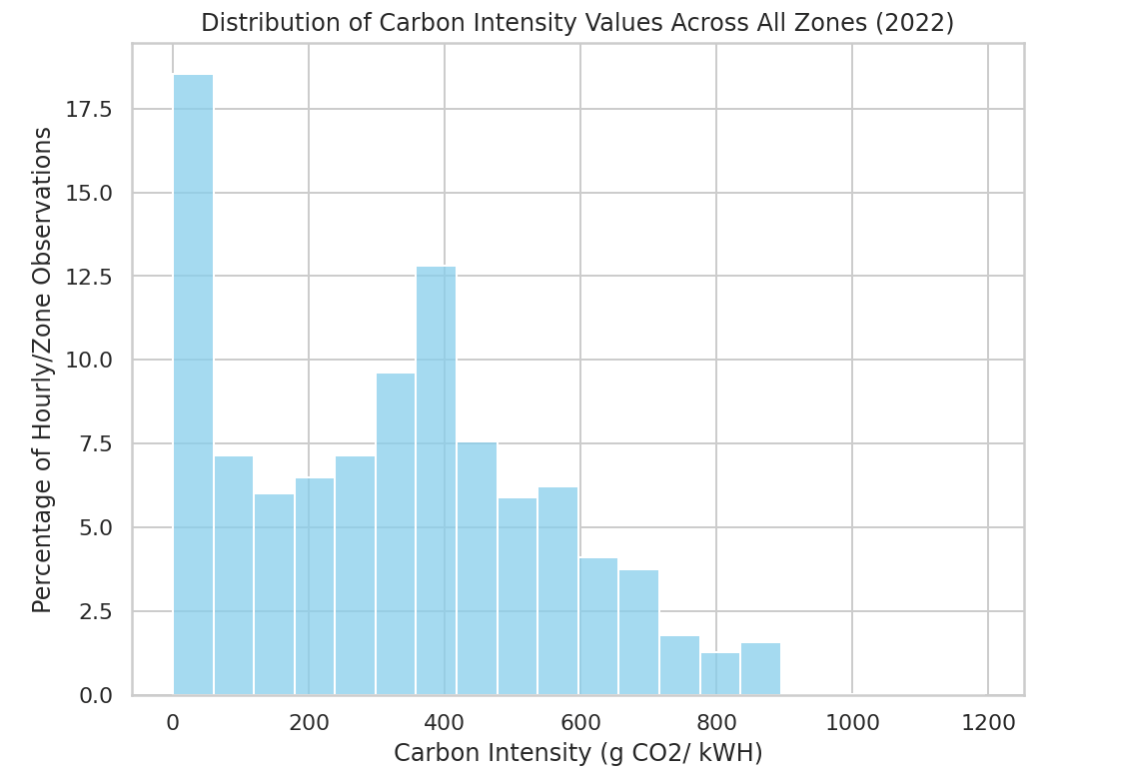}
\caption{2022 hourly carbon intensity distribution for all zones from Electricity Maps (used in Google's carbon emissions estimates).}
\label{fig:carbon_intensity_distribution}
\end{figure}

Figure \ref{fig:carbon_intensity_distribution} shows the distribution of carbon intensity values (gCO$_{2}$e/kWh) for all Electricity Maps zones across all hours for the year 2022 . The global mean is 320.8 gCO$_{2}$e/kWh with a standard deviation of 227.5 gCO$_{2}$e/kWh. It is important to note that this distribution does not represent Google's energy carbon intensity, as our energy consumption is  not equally distributed across all Electricity Maps zones.

Electricity Maps estimates the carbon intensity of electricity in a particular location using a consumption-based accounting method \citep{tranberg2019real}, which builds on flow tracing techniques. The method follows power flow on the transmission network mapping the paths between the location of generation and the location of consumption \citep{tranberg2019real}. For example, the methodology shows that Denmark has a lot of coal production, but during many hours it imports substantial amounts of hydro and wind energy from Norway and Sweden. In each hour, the method traces power flows to estimate the generation mix that is used to supply the power that Denmark consumes, which includes Denmark's coal generation, but also the imported wind and hydro from neighboring grids. Then, the methodology uses the Intergovernmental Panel on Climate Change's emissions factors to calculate the carbon intensity associated with that generation mix.   

After hourly energy consumption (kWh) is allocated to each internal service in each data center cluster, we use PUE data and data from Electricity Maps to estimate the corresponding carbon emissions (kg CO$_{2}$e) associated with the electricity consumed. To do so we link each data center cluster with the Electricity Maps zone it resides in. All data center clusters are embedded within, at-most, a single Electricity Maps zone. Then, we multiply power generated per hour per internal service per data center cluster by its corresponding carbon intensity value per hour per zone. For locations that do not have real-time carbon intensity data, we use the International Energy Agency (IEA) electricity grid annual averages.

More formally:
\begin{equation}
CF_{j, C, h} = (P_{j, C, h}) * (PUE_{C, h}) * (CI_{Z(C), h})
\end{equation}

Where:
\begin{itemize}
    \item  $CF_{j, C, h}$ is the carbon footprint for user $j$ in cluster $C$ in hour $h$, in kg CO$_{2}$e. 
    \item $P_{j, C, h}$ is the IT energy consumption of user $j$ in cluster $C$ in hour $h$, in MWh. This is calculated according to the approach described in Section \ref{sec: power allocation}. Note that in that section we omit the $h$ for simplicity; the calculation steps are the same in any hour $h$
    \item $PUE_{j, C, h}$ is the PUE of cluster $C$ in hour $h$. This is the ratio of total data center energy consumption divided by IT energy consumption. 
    \item $CI_{z, h}$ is the carbon intensity value in electricity zone $z$ in hour $h$, in kgCO$_{2}$e/MWh. The map $Z(C)$ tells us which specific electricity zone each physical cluster $C$ is located in.
\end{itemize}
  
Note that these carbon emissions are based on the base carbon intensity of the corresponding electricity grid; they do not account for Google's carbon-free energy purchases in a particular region. This means that the carbon emissions we currently allocate to customers are based on the \emph{location-based} accounting standard. We are currently working on an update to also provide customers with a view of \emph{market-based} emissions that account for Google's renewable-energy purchases, since this is one important way we can reduce data center emissions. For Google Cloud, from January - December 2023, on average, market-based emissions was 63\% less than location-based emissions, with the ratio of location-based to market-based emissions varying by Google Cloud location and time of usage.

\subsection{Allocating carbon emissions to customers} \label{customer_allocation}
Once we have data on electricity consumption and carbon emissions per internal user, hour, and data center campus, Google allocates electricity and carbon emissions from a Google Cloud product to its multiple products (SKUs) and to external customers. We illustrate this in Figure \ref{fig:stage_2_diagram}:

\begin{figure}[h]
\centering
\includegraphics[scale=0.55]{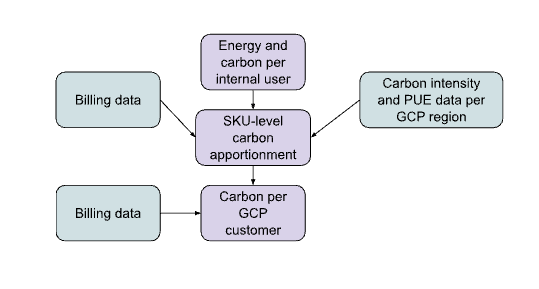}
\caption{High-level overview of allocation of carbon from internal users that provide the resources for Google Cloud products to external Google Cloud customers.}
\label{fig:stage_2_diagram}
\end{figure}

Each SKU (``stock keeping unit'') is mapped to a Google Cloud product, and we connect each Google Cloud product to an internal user that provides the resources for that product. We then estimate the hourly energy for each service in each location. Next, we use revenue data to allocate energy proportional to list prices for different SKUs within a product. Finally, we add carbon intensity information per Google Cloud product and Cloud region. 

\subsubsection{Using Revenue Data to Allocate Electricity to SKUs within Internal Services}

For each product, we conduct the following procedure. Let $T$ be a Google Cloud product offered by user $j$, such as Compute Engine or BigQuery. User $j$ might offer multiple Google Cloud products. Furthermore, let $K_j$ be the set of SKUs offered by user $j$ across all of its Google Cloud products.

\begin{itemize}
    \item Calculate the total global electricity consumption for the internal user, $P_j = \sum_{C,h} P_{j,C,h}$.
    \item Calculate total usage for each SKU, $U_{s}$, for each SKU offered by the internal user $j$.\footnote{Within each service are a list of SKUs, which are distinct offerings for which customers are charged. Examples include BigQuery BI Engine and BigQuery Storage API, both of which are available to purchase for the BigQuery Service.} Each SKU has a specific unit to measure its usage. Examples include vCPU*hour for Compute Engine SKUs or GiB*month for Cloud Storage SKUs. The total usage $U_{s}$ represents the sum of usage from all users, in all regions, for the same time period over which $P_j$ is calculated.
    \item Get the \textbf{list} price for each SKU, $C_{s}$. We do not include Commitment SKUs, Committed Use Discounts, or Sustained Use Discounts in our measurement of $C_{s}$, as we only want to consider list price to fairly allocate energy to end-customers.
    \item  We calculate energy per usage unit for each SKU by a formula which ensures that the energy consumption/usage unit is proportional across SKUs to price/usage unit, while also ensuring that the total energy consumption sums appropriately to the total energy consumption for the internal service provider. The equation for calculating allocated energy per usage unit for a SKU, $X_{s}$ is:
    
    $$ X_{s} = P_j \frac{1}{U_s}\frac{U_s * C_{s}}{\sum_{v \in K_j} U_{v} * C_{v}} = \frac{P_j C_s}{\sum_{v \in K_j} U_{v} * C_{v}}$$
    
    In this way, the energy usage sums appropriately, i.e. $ \sum_{s \in K_j} U_{s} * X_{s} = P_j$. Additionally, energy per usage unit is proportional to list prices for any pair of SKUs, $\frac{X_{s}}{X_{s'}} = \frac{C_{s}}{C_{s'}}$ for any $s$, $s'$, provided by the same service (i.e. Compute Engine).
\end{itemize}

Consider a simple example below: Google Cloud only offers two SKUs for purchase, SKU A and SKU B. Assume that 1 unit of SKU A costs \$28 and 1 unit of SKU B costs \$16 (i.e. one unit of SKU A costs 1.75 times more than one unit of SKU B). Our methodology allocates 1.75 times more energy per unit of usage to SKU A than to SKU B. From the usage and cost data, we have the \textcolor{blue}{blue} data in Table \ref{table: sku energy}. We calculate the numbers in \textcolor{red} {red} cells, first adjusting the total energy estimates to align with the overall energy consumption (30 Wh), and then calculating the final adjusted energy/unit figures.  

\begin{table}[H]
\centering
\begin{tabular}{|c|c|c|c|c|}
\hline
\multirow{2}*{SKU} & Quantity & \multirow{2}*{Normalized Cost} & \multirow{2}*{Est. Energy} & Est. Energy / \\
& (SKU Units) & & & SKU Unit \\
\hline
A & \textcolor{blue}{15} & \textcolor{blue}{1.75} & \textcolor{red}{16 Wh} &  \textcolor{red}{1.62 Wh} \\
\hline
B& \textcolor{blue}{10} & \textcolor{blue}{1} & \textcolor{red}{14 Wh} &  \textcolor{red}{0.92 Wh} \\
\hline
Total &  &  & \textcolor{blue} {30 Wh} &  \\
\hline
\end{tabular}
\caption{Example of allocation of energy from Google Cloud products to SKUs.}
\label{table: sku energy}
\end{table}

The methodology solves for X = 0.92 from the following equation: $10 * 1.75 * X + 15 * 1 * X = 30$, and the ratio of estimated energy/unit across the two SKUs is 1.75 (1.62/0.92), which is equal to the ratio of the list price between the two SKUs. One limitation of this approach is that customers do not receive a signal about how switching from SKU A to SKU B can reduce their actual emissions, but rather their allocated emissions based on list prices. For example, if SKU A and SKU B actually consume the same amount of electricity and are in the same Google Cloud region, SKU A might have a 20\% lower allocated carbon footprint than SKU B, just because SKU A has a 20\% lower list price than SKU B. This discrepancy does not inform the customer about the actual relative footprint of SKU A and SKU B. The methodology was designed such that we could use billing data to apply the same methodology across all Google Cloud products, but future work could consider physical differences in the service stack, such as use of GCU, RAM, SSD, and HHD resources at the SKU-level, along with prices as factors for allocating emissions. 

\subsubsection{Google Cloud Products' Carbon Footprint Per Cloud Region}

Many Cloud products and specific SKUs are offered in multiple Cloud Regions. We estimate emissions of each SKU and service per region because this is the level at which Google Cloud customers can design their Cloud architecture. If customers are interested in updating their Cloud architecture to optimize for lowering their location-based emissions, one of the most effective ways to do this is to move their workload for a given service to a lower-carbon region.

\textit{How do we handle resource usage in different locations than customer workloads?} Some Cloud products use resources in places that are not exactly equivalent to their customers' footprint. For example, a product might run back-end analysis or testing in a different cluster with a lower carbon footprint than their customers' workloads. The design decisions we made are meant to address this. First, we allocate energy globally and adjust for carbon intensity by region, rather than allocating carbon emissions directly by region. This accounts for the fact that a service might have overhead resources or supporting resources in just one location--we want to make sure that overhead energy consumption gets allocated to all customers, not just those in the particular region where the overhead resources are located. Second, we use an adjustment term to capture overhead to ensure that the total carbon emissions of the service are allocated to users.

We estimate emissions of each SKU per Cloud Region $r$ according to the following protocol. First, for any set of Google Cloud products provided by internal Google user $j$, we estimate the net carbon intensity (accounting also for differences in PUE and grid carbon intensity) of the product's offerings in each Cloud Region, $$C_{j,r} = \frac{\sum_{h, C \in r} CF_{j,C,h}}{\sum_{h, C \in r} P_{j,C,h}}$$. 

A natural assumption is that the carbon emissions per usage unit of any SKU $s$ offered in region $r$ is $C_{j,r} X_s$. We make a slight adjustment to ensure that the total carbon emissions allocated across all SKUs adds up to the total internally measured carbon footprint. Thus, let $\overline{C}_{s,r}$ be the carbon intensity per usage unit of a SKU $s$ that is offered by internal user $j$ in region $r$.

Let $U_{s,r}$ be the usage of SKU $s$ in the particular region $r$. First, we calculate the total global electricity consumption for the internal user, $CF_j = \sum_{C,h} CF_{j,C,h}$. Then, to find $\overline{C}_{s,r}$, we solve a ``carbon balance'' equation:
$$ CF_j = \sum_{r, s \in K_j} \overline{C}_{s,r} X_s U_{s,r} = \sum_{r, s \in K_j} \alpha C_{j,r} X_s U_{s,r} $$ 
by setting $\alpha = CF_j / (\sum_{r, s \in K_j} C_{j,r} X_s U_{s,r} )$, and therefore $\overline{C}_{s,r} = \alpha C_{j,r}$. The variable $\alpha$ just serves as a re-weighting term to ensure that the total carbon footprint attributed to internal user $j$ gets fully allocated to the SKUs it has provided to customers. This is done because some Cloud products use resources in places that are not exactly equivalent to their customers’ footprint.  Since we employ a carbon accounting methodology, the $\alpha$ terms ensures that the total carbon footprint that each internal user $j$ is responsible for (Compute Engine for example) equals the total carbon footprint that are reported to Cloud customers.

\subsubsection{Final Allocation to End Customers}
Finally, we calculate the emissions allocated to each Google Cloud billing account $b$ for each external Google Cloud Product in each Google Could region. The following is done on a \textbf{monthly} basis. As before, let $T$ be a particular Google Cloud product. Let $K_T$ be the set of SKUs included in product $T$. Let $U_{s,r,b}$ be the usage of SKU $s$ in region $r$ measured for Google Cloud billing account $b$ on the particular day. Then, customer $b$'s allocated carbon footprint for product $T$ in region $r$ is given by 

$$
CF_{b,T,r} = \sum_{s \in K_T} \beta \overline{C}_{s,r} X_s U_{s,r,b}.
$$

Similarly to how we've defined $\alpha$, $\beta$ is an overhead term to ensure all of Google Cloud's emissions get allocated to customers. Specifically, $\beta = (\sum_{j} CF_j )/(\sum_{s \in T, r} \overline{C}_{s,r} X_s U_{s,r}).$ This allows us to ensure that any overhead that's not directly allocated to customers (e.g. machines that are held by a generic internal user for other Cloud teams to use as necessary) is allocated to customers.

 
This gives us an estimate of total carbon emissions per billing account, broken down by location and by Google Cloud product. We present end-customers with an estimate of their total emissions per-product, per-region, per-month.

\section{Future work and discussion}
We developed this approach with scale in mind: it was essential to provide a fair, reasonable, Google-wide allocation of carbon emissions to all Google teams, products, and customers. This approach necessitated some important trade-offs. In particular, the approach we developed is an allocation-based accounting approach. We could improve aspects of the existing allocation, or consider more substantial departures from the existing allocation method to produce causal estimates of carbon emissions. Here are some examples of potential improvements: 
\begin{itemize}
    \item \reva{For example, for the customer-allocation decisions that currently use economic factors, like cost or revenue (e.g. those described in Section \ref{customer_allocation}), we could improve our data collection to use physical allocation instead, replacing revenue-based allocations with more specific knowledge of physical resource requirements and their energy use.}
    \item \revb{We could improve the assumptions we use in physical allocations. For example, we currently use GCU usage to allocate dynamic power usage when multiple users share a single machine. In reality, users with above-average memory usage are probably responsible for more energy consumption. We could test the extent to which memory usage impacts overall machine energy consumption and augment the model to consider GCU usage and memory usage simultaneously to allocate dynamic power on shared machines.}
    \item We could try to derive a more directly causal estimate of carbon emissions. In some cases, estimating causal impacts could complement the accounting and reporting-focused approach we outlined here.
    \item We could build validation tools based on direct measurement of energy consumption of specific SKUs, when possible. This would allow us to directly validate potential improvements to allocation methodologies by comparing their accuracy for some SKUs. 
\end{itemize}
The allocations presented here are a useful starting point, but we know that future improvements in any of these directions will make it easier to identify and suggest software optimization to reduce carbon emissions. 

The work described here focuses on Scope 2 carbon emissions--those arising from purchased energy (electricity) in data centers. As described in \citep{gupta2021chasing}, as electricity grids in data center locations become cleaner, and as Cloud providers like Google continue to buy more hourly-matched carbon-free energy, the electricity-related emissions associated with a unit of data center compute will continue to shrink. As that happens, the manufacturing emissions associated with data center hardware will become an increasingly important segment of data center emissions. Future work should focus on more accurately measuring data center hardware manufacturing emissions and allocating those emissions to internal users and external Cloud customers. These improvements will help spread the benefits of accurate accounting and measurement to an increasingly critical segment of data center emissions. 

Future work will also add new data to enable, first, enhanced reporting using the \emph{market-based} emissions standard, and, later, improved data to enable carbon-reduction efforts. First, we will launch market-based emissions reporting to Cloud customers, which accounts for Google's carbon-free energy procurement by matching Google's procurement to relevant data centers loads according to GHGP standards. While the Cloud Carbon footprint location-based metrics use \textit{hourly} emissions factors, market-based metrics will use \textit{annual} greenhouse gas emissions factors to be consistent with Google's company-wide annual Scope 2 emissions accounting methodology. For this reason, we anticipate that customers will use market-based metrics primarily for \textit{reporting purposes}. In the longer term, we hope to create more tools to help customers optimize their Google Cloud architecture to reduce electricity consumption and carbon emissions. We hope to build metrics and measurement tools, based on customer-specific use cases, that are most helpful for directing actions to reduce carbon emissions. 



\bibliographystyle{plain}
\bibliography{references}
\end{document}